# The Nuclear Cycle that Powers the Stars: Fusion, Gravitational Collapse, and Dissociation

O. Manuel[1], Michael Mozina[2], and Hilton Ratcliffe[3]

The finding of an unexpectedly large source of energy from repulsive interactions between neutrons in the 2,850 known nuclides has challenged the assumption that H-fusion is the main source of energy that powers the Sun and other stars. Neutron repulsion in compact objects produced by the collapse of stars and collisions between galaxies may power more energetic cosmological events (quasars, gamma ray bursts, and active galactic centers) that had been attributed to black holes before neutron repulsion was recognized. On a cosmological scale, nuclear matter cycles between fusion, gravitational collapse, and dissociation (including neutron-emission) rather than evolve in one direction by fusion. The similarity Bohr noted between atomic and planetary structures may extend to a similarity between nuclear and stellar structures.

[1] Nuclear Chemistry, University of Missouri, Rolla, MO 65401 USA, om@umr.edu

[2] President, Emerging Technologies, P. O. Box 1539, Mt. Shasta, CA 96067, USA, michael@etwebsite.com
 1-800-729-4445

[3] Astronomical Society of South Africa, PO Box 9, Observatory 7935, SOUTH AFRICA, ratcliff@iafrica.com

I.   INTRODUCTION

Hydrogen (H) and other lightweight elements are dominant on stellar surfaces and in the interstellar medium. Since the classical 1957 paper on element synthesis in stars by Burbidge *et al.* [1] (hereafter B2FH), it has been widely assumed that H-fusion is the main driving force for stellar luminosity and ordinary stellar evolution. The idea of a universe driven in one direction by H-fusion fits with the concept of H-production in an initial "Big Bang". However, a recent analysis of the systematic properties of all 2,850 known nuclides [2] revealed an even larger source of energy from repulsive interactions between neutrons in condensed nuclear matter [3-5].

Those results [3-5] and the abundances of isotopes and elements in meteorites, planets, the solar wind, the solar photosphere, and solar flares [6-9] show that:

a) The Sun and other stars act as plasma diffusers, sorting lighter atoms to their surfaces.

b) The interior of the Sun is made of common elements in rocky planets and meteorites – Fe, Ni, O, Si, and S – although the lightest elements (H and He) cover its surface.

c) Neutron-emission from the solar core, a neutron star, is the first step in a series of reactions that has steadily generated luminosity, neutrinos, solar mass fractionation, and an out-pouring of solar-wind hydrogen from the Sun over the past 4-5 Gy.

The Sun is an ordinary star, likely powered by the same processes as other stars. Prior to these recent papers [3-9], compact nuclear matter or black holes had been considered as likely energy sources for short-term energetic events, like gamma ray bursts and quasars, but not as an energy source that might sustain ordinary stars for billions of years. In hindsight, the layers of ordinary atomic matter [10] that separate and insulate the surface of the Sun from its energetic core likely dampen short-term energy fluctuations in the core of the Sun.

Our latest paper [9] includes a few examples of the rigid, iron-rich structures that Mozina [10] noticed below the Sun's fluid photosphere in images from the SOHO and TRACE satellites. Helio-seismology data have since confirmed stratification at a relatively shallow depth beneath the visible photosphere, at ≈ 0.5% solar radii (≈0.005 $R_o$) [11].

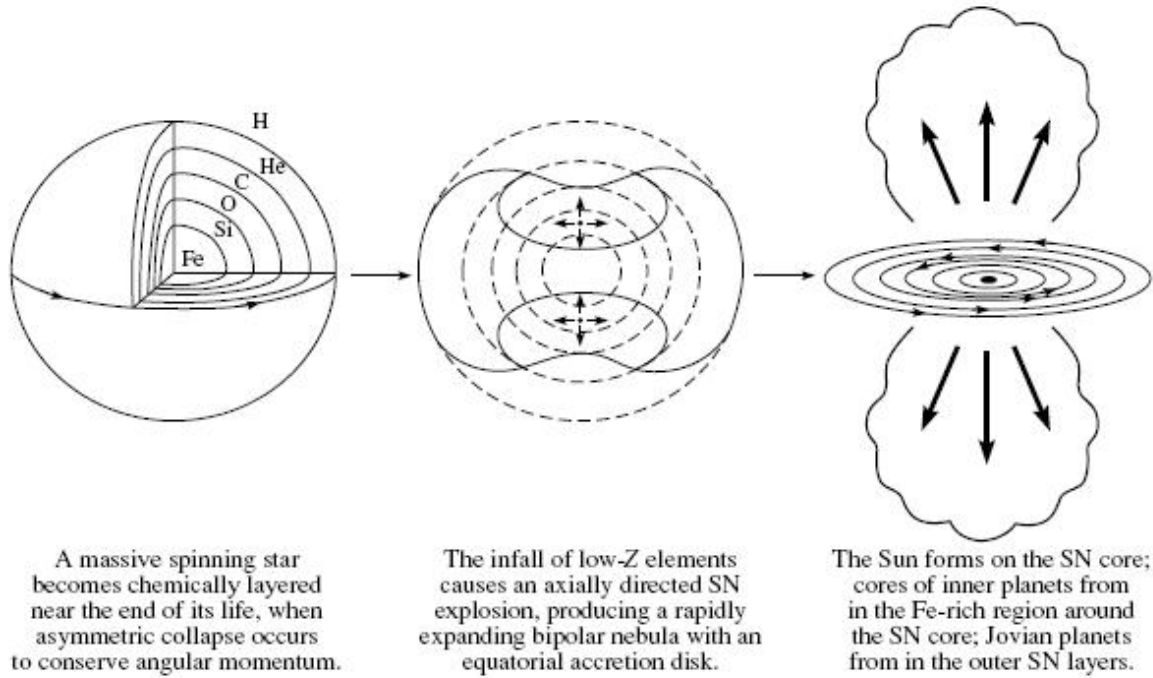

A massive spinning star becomes chemically layered near the end of its life, when asymmetric collapse occurs to conserve angular momentum.

The infall of low-Z elements causes an axially directed SN explosion, producing a rapidly expanding bipolar nebula with an equatorial accretion disk.

The Sun forms on the SN core; cores of inner planets from in the Fe-rich region around the SN core; Jovian planets from in the outer SN layers.

**Fig. 1**. A scenario proposed in the mid-1970s to explain the unexpected link observed between specific isotopes of heavy elements with light element abundances in meteorites at the birth of the solar system [12, 13]. According to this view, the Sun is iron-rich and formed on the collapsed core of a supernova (SN), material near the SN core formed iron cores of the planets near the Sun, and light elements from the outer SN layers formed the giant Jovian planets.

These new findings [9-11] lend credence to **a**.) the suggestion in the mid-1970s that the link of light element abundances with those of specific isotopes of heavy elements in meteorites indicates that the Sun formed on the collapsed core of the supernova that gave birth to the entire solar system [12, 13], as shown in Fig. 1; **b**.) Mozina's conclusion [10] of solar stratification and high levels of electrical and magnetic activity in the Sun's iron-rich stratified layers, **c**.) the suggestion that superfluidity of material in the interior of the Sun causes solar eruptions and climate changes [14], and **d**.) Birkeland's finding [15] at the start of the 20[th] century that many solar features resemble traits observed in the laboratory on magnetized metal spheres, including the link of the *aurora borealis* to solar magnetic activity.

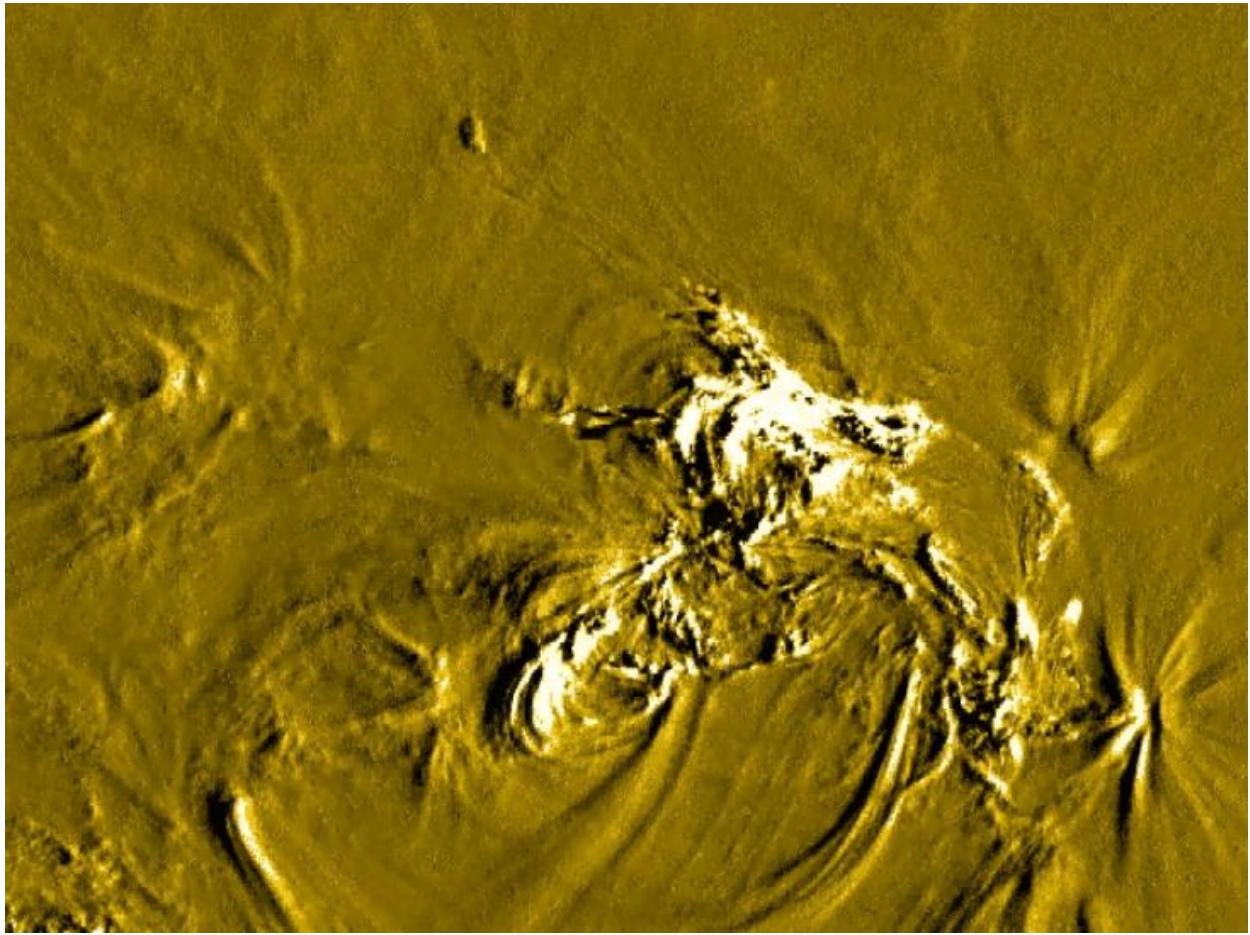

**Fig. 2.** A "running difference" image of the rigid, iron-rich structures beneath the photosphere in a small part of the Sun's surface revealed by the TRACE satellite using a 171 Å filter [10]. This filter is specifically sensitive to light emitted from Fe (IX) and Fe (X) iron ions. A movie made by the Lockheed Martin's TRACE satellite team shows a solar flare and mass ejection (moving towards the upper left of the image) from this Active Region 9143 in 171Å light on 28 August 2000. The movie is here: http://trace.lmsal.com/POD/movies/T171_000828.avi, or it is available here: http://vestige.lmsal.com/TRACE/Public/Gallery/Images/movies/T171_000828.avi

Fig. 2 shows rigid structures in a "running difference" image of one small part of the Sun's iron-rich substructure, Active Region 9143, on 28 August 2000. The TRACE satellite generated this image using a 171 Å filter that is specifically sensitive to light emitted by Fe (IX) and Fe (X) iron ions. The team that operates the TRACE satellite system for NASA made a movie of the flare and mass ejection event that occurred from this Active Region 9143 on 28 August 2000. You can see the movie at http://trace.lmsal.com/POD/movies/T171_000828.avi or it is available here: http://vestige.lmsal.com/TRACE/Public/Gallery/Images/movies/T171_000828.avi.

For the present paper, chemical stratification and electromagnetic active near the surface of the Sun [3-11] are only of interest in demonstrating that energy sources other than H-fusion may operate deep in its interior. Despite well-documented evidence of extensive electrical and magnetic activity near the solar surface [10, 14-15], we are more concerned here with the nuclear forces that have been able to sustain luminosity and an outpouring of hydrogen from the surface of the Sun and other stars over cosmological time scales of billions of years.

One external feature of the Sun, solar-induced variations in the geomagnetic field with a 2.65 h oscillation period, provided a hint almost three decades ago that the Sun itself might be a pulsar [16]. We will show below that repulsive interactions between neutrons in such condensed nuclear matter is the driving force that generates an outpouring of hydrogen and luminosity from a chemically stratified, iron-rich Sun [3-11] and a likely energy source elsewhere in the cosmos.

## II.     THE ENERGY SORCE FOR AN IRON-RICH, STRATIFIED SUN

A systematic enrichment of the lightweight isotopes of all five stable noble gases was recognized in the solar wind in 1983, extending over half of the entire mass range of the stable elements from A = 3 amu to 136 amu (atomic mass units) [6]. Other measurements [7-9] independently confirmed that the Sun selectively moves lighter ions into the photosphere, over the mass range of A = 25-207 amu [8], leaving little doubt that the interior of the Sun is iron-rich [6] like the material that formed iron meteorites and iron cores of rocky planets at the birth of the solar system [12-13]. Iron is however made of tightly packed nucleons [2] and is therefore an unlikely source of nuclear energy. This impasse lasted several years before it was realized in 2000 that repulsive interactions between neutrons in the solar core might be the source of both solar luminosity and the outpouring of solar-wind hydrogen from the surface of the Sun [3-5].

In the spring of 2000 five graduate students– Cynthia Bolon, Shelonda Finch, Daniel Ragland, Matthew Seelke, and Bing Zhang –who were enrolled in a graduate class entitled, *"Advanced Nuclear Chemistry: A Study of the Production and Decay of Nuclei"*, worked with the instructor, O. Manuel, to see if the properties of the 2,850 known nuclides [2] might reveal an unrecognized source of nuclear energy. Fig. 3 is a pictorial summary of the evidence they uncovered for repulsive interactions between neutrons in the nuclei of ordinary nuclear matter.

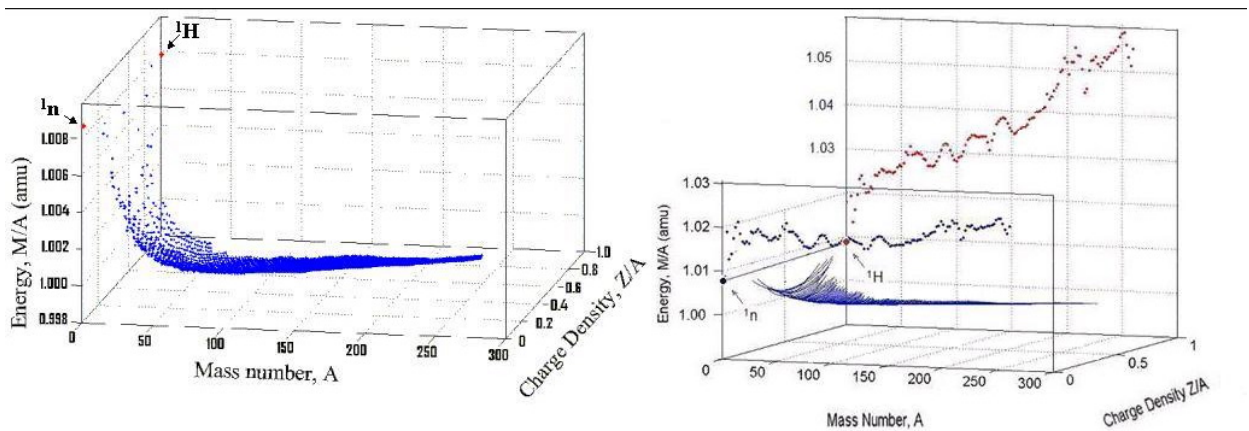

**Fig. 3.** <u>Left</u>: The "Cradle of the Nuclides" is revealed when properties of ground-state nuclides [2] are plotted on a 3-D plot of mass per nucleon, M/A, *versus* charge density, Z/A, *versus* mass number, A. <u>Right</u>: For all values of A > 1 amu, mass parabolas defined by the data at other mass numbers intercept the front plane at (Z/A = 0, M/A = (M/A)$_{neutron}$ + ≈ 10 MeV) [3-5].

Data points in the drawing on the left side of Fig. 3 are experimental, but those on the right side at Z/A = 0 and Z/A = 1 are experimental only at A = 1, where they represent the neutron ($^1$n) and the lightest hydrogen isotope ($^1$H), respectively. The other data points on the right side of Fig. 3 were calculated from the mass parabolas defined by the mass data [2] at each value of A>1. Except for Coulomb repulsion, the n-n and p-p interactions are symmetric [3]. Values of the potential energy per nucleon (M/A) at Z/A = 0 and Z/A = 1 are therefore similar at low A.

As the value of A increases, values of M/A at Z/A = 1 become increasingly larger than those at Z/A = 0 because of the increasing contribution from Coulomb repulsion between positive charges. Thus repulsive interactions between protons prevents the formation of massive stellar objects compressed to nuclear density at Z/A = 1.

However at Z/A = 0 repulsive interactions between neutrons, documented by high values of M/A on the right side of Fig. 3, generate solar luminosity, energy in neutron stars, and an outpouring of hydrogen from the surfaces of the Sun and ordinary stars [3-5, 7-9]. In the cosmos,

**(neutron-emission) + (neutron-decay) => (hydrogen production),**

despite the accepted view that neutron stars are "dead" nuclear matter with each neutron in a neutron star having about 93 MeV less energy than a free neutron [17].

On the mass scale of ordinary nuclear matter, i.e., for A ≤ 300 amu, repulsive interactions between neutrons may contribute to the following observations:

**a.)** Spontaneous neutron-emission from neutron-rich nuclei over essentially the entire mass range of nuclear matter, e.g., $^5$He – $^{149}$La [2];

**b.)** The rhythmic cycle in values of the potential energy per nucleon at Z/A = 0 (Fig. 3) over the entire mass range (A) caused by geometrical changes in the packing of nucleons [4];

**c.)** Spontaneous fission of heavy nuclei with A ≥ 230 amu [2]; and

**d.)** A halo cloud of neutrons extending beyond the charge radius in neutron-rich light nuclei, e.g., the two-neutron cloud at the surface of $^{11}$Li outside the core nucleus of $^9$Li [18].

Coulomb repulsion becomes increasingly important in heavy nuclei, and neutrons are likely to be concentrated at the core rather than the surface of these. This transition in the internal structure of ordinary nuclei seems to occur near A ≈ 150 amu, where neutron-emission ceases and alpha-emission begins [2]. For A >150 amu, the potential energy per nucleon (M/A) at Z/A = 0 (See right side of Fig. 3) gradually begins to increase with A [5]. This trend was used to extrapolate an upper limit of ≈ 22MeV on the excitation energy of a neutron in a neutron star [5].

The asymmetric fission of heavy nuclei into heavy and light mass fragments with a mass ratio of ≈ 1.6/1 has long been linked to the closed shells of neutrons at N = 82 and 50, respectively. This empirical observation suggests that neutron repulsion in the core of heavy nuclei may contribute to nuclear fission.

Properties a.)–c.) of ordinary nuclear matter may also occur on the cosmological mass scale of dense, neutron-rich stellar objects for A > ≈ $10^{57}$ amu ≈ 1 solar mass ($M_o$). I.e., the similarity Bohr [19] noted between atomic and planetary structures may extend to a similarity between nuclear and stellar structures. However, another force comes into play on the cosmological mass scale that is unimportant on the mass scale of ordinary nuclear matter shown in Fig 3 – gravity.

## III. THE NUCLEAR CYCLE THAT POWERS THE COSMOS

It had long been assumed that gravitational collapse produces the neutron stars that are seen near the center of supernova (SN) debris. That is the scenario outlined in Fig 1 to explain the presence of a neutron star at the core of the Sun. There is indeed compelling evidence that highly radioactive debris of a supernova that exploded here 5 Gy ago formed the solar system before isotopes and elements from different SN layers had completely mixed [9, 12-13, 20-21].

However there are also indications that the elements in the parent star (Fig. 1) had also been sorted by mass [9]. This suggests the presence of a carrier gas, an upward flow of $H^+$ ions generated by neutron-emission and neutron-decay from a neutron star at the core of the parent star, prior to the SN explosion that gave birth to the solar system. Thus, the occurrence of a neutron star in the core of the Sun, in its precursor, and in other ordinary stars [3-9] implies that:

- **a.)** Stellar explosions may expose, but do not necessarily produce, the neutron stars that are seen in stellar debris; and
- **b.)** Neutron stars at the centers of ordinary stars were not produced one-at-a-time in SN explosions but were more abundantly made in higher energy events, such as galactic collisions that likely altered or produced the current Milky Way.

A recent review on galactic collisions notes that "transient galaxy dynamics", the recurrent collisions and mergers of galaxies, has replaced the classical view that galactic structures formed early in the universe and were followed by slow stellar evolution and the steady build-up of heavy elements [22]. Collisions or mergers of galaxies are highly prevalent, with ~1 in 10 of known galaxies engaged in some stage of physical interaction with another galaxy, and nearly all cohesively-formed galaxies, especially spirals, having experienced at least one collision in their lifetime. *"Galactic collisions involve a tremendous amount of energy. . . . . the collision energy is of order $10^{53}$ J. This is equivalent to about $10^{8-9}$ supernovae, . . ."* [ref. 22, p. 6]

Collisions are highly disruptive to all components of the galaxies, including the nucleus, and astronomers observe the collisional energy in many puzzling forms - quasars, gamma ray bursts, and active galactic centers (AGN). The extreme turbulence of active galactic nuclei (AGN) suggests the interactive presence of massive gravitational concentrations, possibly black holes [23] or super-massive neutron stars that fragment explosively [24] into the multiple neutron stars that then serve as formation sites of new stars. Struck notes in the abstract of his review paper that *"Galactic collisions may trigger the formation of a large fraction of all the stars ever formed, and play a key role in fueling active galactic nuclei"* [ref. 22, p. 1].

Matter is ejected from the massive object in the galaxy core in the form of jets, perhaps caused by Bose-Einstein condensation of iron-rich, zero-spin material into a super-fluid, super-conductor [25] or by super-fluidity and quantized vortices in the central neutron stars [26].

Hubble Space Telescope observations confirmed a hierarchical link suggested by Arp [27] between collisional systems and quasi-stellar objects - quasars. Quasars are frequently seen grouped, in pairs or more, across active galaxies, and are physically linked to the central galaxy by matter bridges. Isophote patterns indicate that the direction of motion of the quasars is *away* from their host galaxy, thereby stretching and weakening the matter bridge until the quasar separates completely. The implication is certain—quasars are physical ejecta from AGN that form the nuclei of nascent galaxies. The review paper states *". . . HST observations.. . . provide direct evidence that some, and the implication that most, of the quasar hosts are collisional systems"* [ref. 22, p. 105].

AGN, quasars, and neutron stars are highly prevalent, observable phenomena in all parts of the known universe. They have two significant properties in common: Exceptionally high specific gravity and the generation of copious amounts of "surplus" energy. In view of the repulsive forces recently identified between neutrons [3-5] and the frequency and products of galactic collisions [22], we conclude that neutron repulsion is the main energy source for the products of galactic collisions.

Fig. 4 is a pictorial summary of the main features of the nuclear cycle that powers the cosmos: 1. Fusion;  2. Gravitational Collapse; and  3. Dissociation (neutron emission).

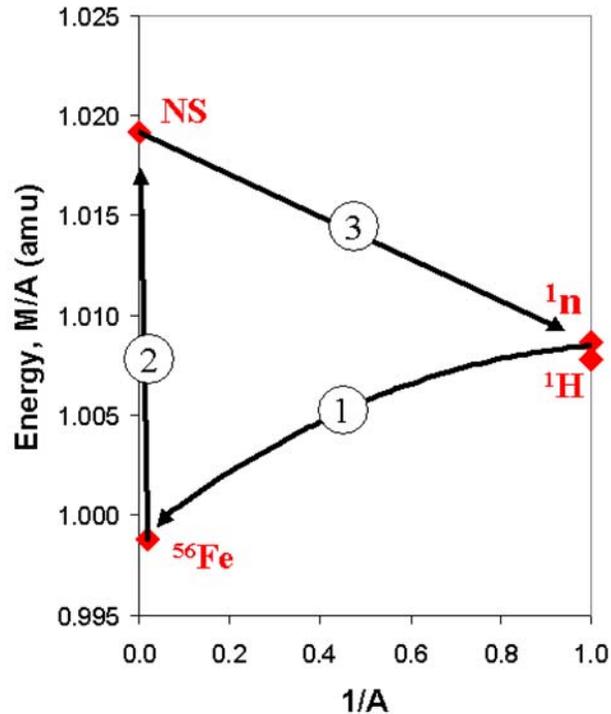

**Fig. 4.** The nuclear cycle that powers the cosmos: **1**. Fusion converts lightweight nuclei, like $^1$H, into heavier ones, like $^{56}$Fe; **2**. Gravitational collapse of ordinary atomic matter makes compact cosmological objects with Z/A = 0; and **3**. Neutron-emission and neutron-decay produce the hydrogen fuel used in step **1**.

## IV.   CONCLUSION

Neutron-rich stellar objects produced by stellar collapse and/or galactic collisions exhibit some of the features that are related to neutron repulsion in ordinary nuclei:

**a.)** Spontaneous neutron-emission from a central neutron star sustains luminosity and the outflow of hydrogen from the Sun and other ordinary stars;

**b.)** As a neutron star ages and loses mass, changes in the potential energy per neutron may cause instabilities due to geometrical changes in the packing of neutrons (See the cyclic changes in values of M/A on the right side of Fig. 3 at Z/A = 0); and

**c.)** Spontaneous fission may fragment super-heavy neutron stars into binaries.

The similarity Bohr noted in 1913 [19] between atomic and planetary structures extends to a similarity between nuclear and stellar structures. The recent finding [28] of a massive neutron star (CXO J164710.2-455216) in the Westerlund 1 star cluster where a black hole was expected observationally reinforces our doubts about the collapse of neutron stars into black holes.

## ACKNOWLEDGEMENTS

Support from the University of Missouri-Rolla and permission from the Foundation for Chemical Research, Inc. (FCR) to reproduce figures from FCR reports are gratefully acknowledged. NASA and Lockheed Martin's TRACE satellite team made possible the image of rigid, iron-rich structures beneath the fluid photosphere and the movie of a solar flare coming from that region (Fig. 2). Students in Advanced Nuclear Chemistry (Chem. 471) in the spring semester of 2000 – Cynthia Bolon, Shelonda Finch, Daniel Ragland, Matthew Seelke and Bing Zhang – contributed to the development of the "Cradle of the Nuclides" (Fig. 3) that exposed repulsive interactions between neutrons. Moral support and encouragement from a few former UMR administrators were critical to the completion of this study: Former UMR Chancellors, Dr. Raymond L. Bisplinghoff (1975-1976) and Dr. Gary Thomas (2000-2005); former UMR Deans of Arts and Sciences, Dr. Marvin M. Barker (1980-1990) and Dr. Russell D. Buhite (1997-2002), and former UMR Chair of Chemistry, Professor Stig E. Friberg (1976-1979). This paper is dedicated to the memory of the late Professor Paul Kazuo Kuroda and his former student, Dr. Dwarka Das Sabu, for their deep personal commitment to the basic precepts of science and for their many discoveries that laid the foundation for the conclusions reached here including references [12, 13, 20, 21].

**REFERENCES**

1. E. M. Burbidge, G. R. Burbidge, W. A. Fowler and F. Hoyle, *Rev. Mod. Phys.,* **29**, 547-650 (1957).

2. J. K. Tuli, *Nuclear Wallet Cards* (National Nuclear Data Center, Brookhaven National Laboratory, Upton, NY, 6$^{th}$ Ed., 74 pp., 2000)

3. O. Manuel, C. Bolon, A. Katragada, and M. Insall, *J. Fusion Energy,* **19**, 93-98 (2000).

4. O. Manuel, E. Miller, and A. Katragada, *J. Fusion Energy,* **20**, 197-201 (2001).

5. O. Manuel, C. Bolon and M. Zhong, *J. Radioanal. Nucl. Chem.,* **252**, 3-7 (2002).

6. O. K. Manuel and G. Hwaung, *Meteoritics,* **18**, 209-222 (1983).

7. O. Manuel and S. Friberg , *in* Huguette Lacoste (Ed), *2002 SOHO 12/GONG + 2002 Proceedings* (*ESA SP-517 SOHO/GONG*, Noordwijk, The Netherlands, pp. 345-348, 2003).

8. O. Manuel, W. A. Myers, Y. Singh, and M. Pleess, *J. Fusion Energy,* **23**, 55-62 (2004).

9. O. Manuel, S. Kamat, and M. Mozina, *in* Jose B. Almeida (Ed), *Proceedings First Crisis in Cosmology Conf.* (AIP, Melville, NY, in press, 2005)   http://arxiv.org/abs/astro-ph/0510001

10. M. Mozina, "The surface of the Sun", http://www.thesurfaceofthesun.com/index.html

11. S. Lefebvre and A. Kosovichev, Ap. *J. Lett*, in press (2005)
    http://xxx.lanl.gov/abs/astro-ph/0510111

12. O. K. Manuel and D. D. Sabu, *Trans. Missouri Acad. Sciences,* **9**, 104-122 (1975).

13. O. K. Manuel and D. D. Sabu, *Science,* **195**, 208-209 (1977).

14. O. K. Manuel, M. W. Ninham and S. E. Friberg, *J. Fusion Energy,* **21**, 193-199 (2002).

15.  K. Birkeland, *The Norwegian Aurora Polaris Expedition, 1902–1903,* pp. 661-678 (1908)
    http://www.catastrophism.com/texts/birkeland/

16. P. Toth, *Nature* **270**, 159-160 (1977).

17. H. Heiselberg, "Neutron Star Masses, Radii and Equation of State", in *Proceedings of the Conference on Compact Stars in the QCD Phase Diagram,* eConf C010815, edited by R. Ouyed and F. Sannion, Copenhagen, Denmark, Nordic Institute for Theoretical Physics (2002) pp. 3-16, http://www.arxiv.org/abs/astro-ph/?0201465

18. W. Nörtershäuser, *Triump Newsletter*, **3**, no. 1 (April 2005). http://www.triumf.info/public/news/newsletter/V3N1/Lithium11.htm

19. N. Bohr, *Phil. Mag,*, **26**, 1-25 (1913).

20. P. K. Kuroda and W. A. Myers, *J. Radioanal. Nucl. Chem.,* **211**, 539-555 (1996).

21. P. K. Kuroda and W. A. Myers, *Radiochimica Acta,* **77**, 15-20 (1997).

22. C. Struck, "Galactic collisions", in *Physics Reports*, **321**, 1-137 (1999) http://arxiv.org/html/astro-ph/9908269/homepage.html

23. M. C. Begelman, R. D. Blandford, and M. J. Rees, *Nature*, **287**, 307-309 (1980).

24. W. K. Brown, *Astrophys. Space Sci.*, **15**, 293-306 (1972).

25. B. W. Ninham, *Physics Lett.,* **4**, 278-279 (1963).

26. P. M. Pizzochero, L. Viverit, and R. A. Brogalia, *Phys. Rev. Lett.* **79**, 3347-3350 (1997).

27. Halton C. Arp, *Catalogue of Discordant Redshift Associations* (C. Roy Keys, Inc, Montreal, Quebec, Canada, 2003) 234pp.

28. M. Muno *et al.*, *Ap. J. Lett.,* in press (2005). See news reports http://chandra.harvard.edu/ http://www.sciencedaily.com/releases/2005/11/051103080649.htm